# Resolution of 2D Currents in Superconductors from a 2D magnetic field measurement by the Method of Regularization

D. M. Feldmann*

*Applied Superconductivity Center, University of Wisconsin – Madison, Madison, Wisconsin, 53706*

The problem of reconstructing a two-dimensional (2D) current distribution in a superconductor from a 2D magnetic field measurement is recognized as a first-kind integral equation and resolved using the method of Regularization. Regularization directly addresses the inherent instability of this inversion problem for non-exact (noisy) data. Performance of the technique is evaluated for different current distributions and for data with varying amounts of added noise. Comparisons are made to other methods, and the present method is demonstrated to achieve a better regularizing (noise filtering) effect while also employing the generalized-cross validation (GCV) method to choose the optimal regularization parameter from the data, without detailed knowledge of the true (and generally unknown) solution. It is also shown that clean, noiseless data is an ineffective test of an inversion algorithm.

## 1. Introduction

Considerable effort has been spent to probe the local critical current density ($J_c$) of high-temperature superconducting (HTS) materials. Of particular interest is YBa$_2$Cu$_3$O$_7$ coated-conductors (CCs) and BiSrCaCuO (BSCCO) tapes, where current percolates and transport $J_c$ values are frequently a macroscopic average of large local variations in $J_c$ (Ref. 1-9). Probing the local $J_c$ may be done directly with transport measurements, but such measurements are destructive and provide information only in localized regions[3,5]. Indirect methods of probing $J_c$, frequently done through a spatially resolved magnetic field measurement, can provide information about the local $J_c$ over large areas[1,2,4,10]. Under certain restrictions, a 2D map of the local $J_c$ in a superconductor can be resolved from a 2D magnetic field measurement through inversion of the Biot-Savart law. The local magnetic field required for the inversion may be obtained through magneto-optical imaging (MOI) or scanning Hall probe techniques[11-15].

This magnetic inverse problem has been addressed many times by a variety of methods[2,16-23], but these methods may suffer from several shortcomings. The inversion of the Biot-Savart law exhibits an inherent instability for non-exact (noisy) data, but the results of these methods are often only presented for clean (noiseless) data, which is a poor test of any method. Experimental data always contains some level of noise, and the performance of any method should be evaluated in the presence of such noise, where the instability of the inversion problem is evident. Several of these methods require a user-chosen parameter, such as the cut-off frequency in the low-pass Fourier filtering method of Roth *et al*[17] or the number of iterations in the conjugate-gradient (CG) method of Wijngaarden *et al*[18,23], but no systematic means of choosing these parameters is presented in those works. While these parameters can be chosen empirically, it is preferable to have a means of choosing such parameters directly from the data. These methods also fail to recognize the inversion of the Biot-Savart law as a member of a larger class of integral equations that have been well studied in the literature. Such shortcomings are overcome in the present work.

Inversion of the Biot-Savart law, separate from the physical representation of reconstruction of current flow, requires the resolution of an integral equation. If all the restrictions required for inversion of the Biot-Savart law are satisfied, then the problem of resolving a 2D current distribution from a 2D magnetic field measurement reduces to an integral equation of the form

$$\int_A K(x-x', y-y') g(x', y') dx' dy' = f(x, y) \tag{1}$$

where the integral kernel $K$ is known, $g$ is to be determined, and $f$ is known at only a discrete number of points and with errors. Equation (1) is a member of a larger class of equations known as *Fredholm Integral Equations of 1$^{st}$ Kind* and is characterized by an inherent instability for non-exact data, since small variations in $f$ can produce large variations in $g$, and $g$ does not depend continuously on $f$ (Ref.24). Such problems are termed *ill-posed*[25]. The degree of ill-posedness of Eq. (1) depends on the form of the kernel $K$, with very smooth kernels generally leading to highly ill-posed problems and $\delta$-function-like kernels being highly desirable. A consequence of this ill-posed nature is that the function $g$ that best satisfies Eq. (1) for a given data set $f$ may deviate greatly from the true solution. First kind integral equations have been well studied in the literature and several methods exist for their evaluation[24,26-35]. One of the most popular of these is the method of Regularization, developed by Phillips[27] in 1962 and expanded by Tihkonov[35] in 1963, which uses *a priori* information about the solution to replace Eq. (1) with a similar, but well-posed problem. For regularization, the a priori information generally concerns the smoothness (or the allowed oscillations) of *g*. Integral equations such as Eq. (1) are not unique to the magnetic inverse problem and occur in many areas of science[36-38]. An excellent primer on First Kind Equations is given by Wing[24]. In this paper, the method of regularization is used to resolve the magnetic inversion problem for both thin film and slab geometry.

## 2. Resolution of the magnetic inversion problem

*2.1. Formulation of the problem*

The geometry of the magnetic inversion problem is shown in Fig. 1. To derive the current flow in a superconductor from a spatially resolved 2D magnetic field measurement $B_z(x,y)$, it is necessary for the current to be adequately approximated as 2D, i.e. that the *z*-component of the current is zero. It is also required that the superconductor be in a magneto-static state such that $\vec{\nabla} \cdot \vec{J} = 0$. This condition can be incorporated by writing the current in terms of the scalar field $g(x,y)$ (Ref. 39),

$$\vec{J} = \vec{\nabla} \times \left( g(x,y) \hat{k} \right). \tag{2}$$

Substituting Equation (2) into the *z*-component of the Biot-Savart law gives

$$B_z(x,y) = \int_{-\infty}^{\infty}\int_{-\infty}^{\infty} K(x-x',y-y')g(x',y')dx'dy' \qquad (3)$$

where the kernel $K(x,y)$ is given by

$$K(x,y) = \frac{\mu_0}{4\pi}\frac{z}{(x^2+y^2+z^2)^{3/2}} \qquad (4)$$

for slab geometry[40],

$$K(x,y) = \frac{\mu_0}{4\pi}\left(\frac{z}{(x^2+y^2+z^2)^{3/2}} - \frac{a+z}{(x^2+y^2+(a+z)^2)^{3/2}}\right) \qquad (5)$$

for thin films of thickness $a$, and

$$K(x,y) = \frac{\mu_0}{4\pi}\frac{2z^2-x^2-y^2}{(x^2+y^2+z^2)^{5/2}} \qquad (6)$$

when the concept of sheet currents is used[39]. $\mu_0$ is the permittivity of free space, $B_z$ is the z–component of the magnetic field (perpendicular to the sample surface), $z$ is the height of the measurement plane above the sample surface, and the $z$ dependence of $K(x,y)$ has been suppressed. In order to determine the current in a sample, Eq. (3) must be resolved for $g(x,y)$. Once $g(x,y)$ has been adequately determined, Eq. (2) can be applied to determine the current vectors $J_x$ and $J_y$. Resolving $g(x,y)$ from Eq. (3) when the data $B_z$ is known only at a discrete number of points and with errors is the main topic of this paper, and is done with the method of Regularization as described in the next section.

*2.2. Regularization*

The method of regularization replaces the problem of inverting Eq. (3) with the problem of minimizing the functional

$$C(g,\lambda) = \left\| \int_{-\infty}^{\infty} \int_{-\infty}^{\infty} K(x-x', y-y') g(x',y') dx' dy' - B_z(x,y) \right\|_2^2 + \lambda\, \Omega[g] \qquad (7)$$

with respect to $g$, where the 2-norm is defined as $\| f(x,y) \|_2^2 = \int_{-\infty}^{\infty} \int_{-\infty}^{\infty} | f(x,y) |^2 dxdy$. The operator $\Omega$ is a user-defined measure of the smoothness of $g$, and $\lambda$ is the *regularization parameter* that controls the trade-off between smoothness and the degree to which Eq. (3) is satisfied. A common (and convenient) choice of $\Omega$ is the norm of an $n^{th}$ derivative of the unknown $g$. It is desirable for the application of Eq. (2) that the first derivatives of $g$ be smooth, so here $\Omega$ is chosen to be

$$\Omega[g] = \left\| \frac{\partial^2 g}{\partial x^2} + \frac{\partial^2 g}{\partial y^2} \right\|_2^2. \qquad (8)$$

The value of Ω[g] will be larger when g is rapidly oscillating (noisy) and smaller when g is smooth. With this choice for Ω it can be shown that the minimizer of Equation (7), $g_\lambda$, is given by[41]

$$g_\lambda(x,y) = \int_{-\infty}^{\infty}\int_{-\infty}^{\infty} \frac{|\hat{K}(u,v)|^2}{|\hat{K}(u,v)|^2 + \lambda(2\pi)^4(u^2+v^2)^2} \left(\frac{\hat{B}_z(u,v)}{\hat{K}(u,v)}\right) e^{i2\pi(ux+vy)} du\,dv \qquad (9)$$

where ^ denotes a Fourier transform. The problem of minimizing $C(g,\lambda)$ has been reduced to a simple Fourier transform with a filter function. However, unlike other Fourier inversion methods[16,17,21], the filter can be directly related back to the imposed smoothness condition on g. Using

$$\hat{h}_{uv} = \sum_{m=0}^{M-1}\sum_{n=0}^{N-1} h_{nm} e^{i2\pi un/N + i2\pi vm/M} \quad \text{and} \quad h_{nm} = \frac{1}{NM}\sum_{v=0}^{M-1}\sum_{u=0}^{N-1} \hat{h}_{uv} e^{-i2\pi un/N - i2\pi vm/M} \qquad (10)$$

as the definitions of the discrete Fourier transform (DFT) and inverse DFT (IDFT) respectively, the minimizer of the discrete version of Eq. (7) is given by

$$g_{nm;\lambda} = \frac{1}{N\Delta_x M\Delta_y} \sum_{v=0}^{M-1}\sum_{u=0}^{N-1} Z_{uv;\lambda} \frac{\hat{B}_{uv;z}}{\hat{K}_{uv}} e^{-i2\pi un/N - i2\pi vm/M} \qquad (11)$$

where the filter $Z_{uv;\lambda}$ is

$$Z_{uv;\lambda} = \frac{\left|\hat{K}_{uv}\right|^2}{\left|\hat{K}_{uv}\right|^2 + 16\lambda \Delta_x^{-2}\Delta_y^{-2}\left(\Delta_x^{-2}\sin^2(\pi u/N) + \Delta_y^{-2}\sin^2(\pi v/M)\right)^2} \; . \tag{12}$$

The second order accurate central difference approximation was used for the discrete version of Eq. (8), assuming a periodic extension of the $\{g_{nm;\lambda}\}$ (Ref. 42). Let the discrete residual norm be defined as

$$\rho(\lambda) = \left\|(K * g_\lambda)_{nm} - B_{nm;z}\right\|_2^2 \tag{13}$$

where the discrete 2-norm is $\left\|f_{ij}\right\|_2^2 = \Delta_x \Delta_y \sum_{i,j} \left|f_{ij}\right|^2$ and the definition of discrete convolution is $(r * s)_{nm} = \Delta_x \Delta_y \sum_{m',n'} r_{(m-m')(n-n')} s_{n'm'}$. The discrete norm $\rho(\lambda)$ is a measure of the degree to which the regularized solution $g_\lambda$ satisfies Eq.(3).

Before Eq. (11) can be applied, it is necessary to choose a value for $\lambda$. A large value of $\lambda$ will result in $g_\lambda$ being quite smooth, with an unnecessary loss of detail. A small value of $\lambda$ will result in the residual norm $\rho(\lambda)$ being small, but the regularized solution $g_\lambda$ may deviate considerably from the true solution. It needs to be emphasized that a small value for the residual norm does not necessarily mean that $g_\lambda$ will be close to the true solution since the data $B_z$ is inexact. The value of $\lambda$ may be chosen empirically by varying $\lambda$ until the smoothness of either the scalar field $g$ or the current vectors $J_x$ and $J_y$ appears most reasonable. This can be rather subjective however, and a more systematic means of choosing the optimal $\lambda$ is desired. Before discussing means of choosing $\lambda$, it is helpful to define what a 'good choice' of $\lambda$ is. The best choice of $\lambda$ is one that minimizes the difference between the approximate and the exact solution as measured in some user-defined way. Here, let the measure be the normalized true mean square error $D(\lambda)$,

$$D(\lambda) = \frac{\|g_{nm;\lambda} - g_{nm;\text{exact}}\|_2^2}{\|g_{nm;\text{exact}}\|_2^2},\qquad(14)$$

where $g_{\text{exact}}$ is the exact (true) solution. If we choose $D(\lambda)$ as the goodness of fit criterion for an approximate solution $g_\lambda$, then the minimizer of $D(\lambda)$, $\lambda_D$, is the best possible choice of $\lambda$ for a given data set $B_z$. More simply, smaller values for $D(\lambda)$ represent better solutions than larger values. In practice, $g_{\text{exact}}$ is generally unknown, and Eq. (14) cannot be minimized directly. In this case, a means of choosing $\lambda$ from the data $B_z$ is desired, such that this choice of $\lambda$ results in a solution close to the exact (unknown) solution as measured by our goodness of fit criterion, $D(\lambda)$. There are multiple methods for choosing $\lambda$ from the data[31,33,43], and one of the most successful is the generalized cross-validation (GCV) method of Wahba[33]. The GCV method is based on statistical considerations, namely, that if an arbitrary element of $B_z$ is left out, then the regularized solution should predict this missing data point well. GCV also seeks to minimize the predictive mean square error.[44] For a more detailed discussion of these points see Ref. 45. Using GCV, the optimal regularization parameter $\lambda_{\text{GCV}}$ is the minimizer of

$$V_{GCV}(\lambda) = \frac{\sum_{v=0}^{M-1}\sum_{u=0}^{N-1}(1-Z_{uv;\lambda})^2 |\hat{B}_{uv;z}|^2}{\left(1 - \frac{1}{MN}\sum_{v=0}^{M-1}\sum_{u=0}^{N-1} Z_{uv;\lambda}\right)} . \quad (15)$$

$V_{GCV}(\lambda)$ is a simple one-dimensional function of $\lambda$ depending only on $\hat{K}$ and $\hat{B}_z$. The calculation of $\hat{K}$ and $\hat{B}_z$ are already required for Eq. (11), and minimization of $V_{GCV}(\lambda)$ is relatively quick.

Once the function $g_\lambda$ has been determined, Eq. (2) still needs to be applied to determine $J_x$ and $J_y$. Since the data $B_z$ contain noise so too will $g_\lambda$, and differentiating a noisy function is itself an ill-posed problem[46]. Small oscillations in $g_\lambda$ can cause large oscillations in its derivatives and therefore in $J_x$ and $J_y$, and the method chosen to take the derivatives will obviously affect the values of $J_x$ and $J_y$. The method chosen to perform the required differentiation in this paper had a slight smoothing effect and proceeds as follows. First, the point of interest ($f_n$) plus a number of data points to the left ($n_L$) and to the right ($n_R$) were fit to a quadratic polynomial $(f_{n-n_L},\ldots,f_{n-1},f_n,f_{n+1},\ldots,f_{n+n_R})$. The estimate of the derivative at the point of interest $f_n$ is then the value of the analytical derivative of the polynomial at that point. Throughout this work, unless stated otherwise, $n_R = n_L = 2$ for a total of $n_R + n_L + 1 = 5$ data points fit to each polynomial, centered on the point of interest. This was carried out in an efficient manner with the use of Savitsky-Golay coefficients[47]. This method of calculating the derivatives results in a slight reduction of spatial resolution. The quadratic polynomials are fit to five grid points, though they would be fully defined by only three. For the examples of this work where noisy data has been used, this reduction in spatial resolution is less than that due to the added noise.

## 3. Numerical Results

### 3.1. L-curve analysis

The regularization functional (Eq. (7)) imposes a trade-off between the smoothness of $g_\lambda$ and the degree to which Eq. (3) is satisfied. This trade-off is shown graphically in Fig. 2, where the smoothing norm $\Omega[g_\lambda]$ is plotted versus $\rho(\lambda)$ for increasing values of $\lambda$. The exact form of the current distribution is given in Fig. 3, and the data $B_z$ have been corrupted with gaussian white noise with variance $\sigma^2 = 0.01 \max\{|B_z|\}$. The smallest value of $\lambda$ occurs in the upper left portion of the plot and the largest in the lower right. It can be seen that small $\lambda$ will result in the norm $\rho(\lambda)$ being small and large $\lambda$ will cause $\Omega[g_\lambda]$ to be small. The optimal value of $\lambda$ as defined by the goodness of fit criterion, $\lambda_D$, is marked with an open circle in Fig. 2. $\lambda_D$ is often in the 'corner' of the '*L*-curve', which gets its name from its 'L' shape. The inset shows the *L*-curve on a linear scale, where the data appears to lie entirely on the plot axes. The corner, or point of maximum curvature of the *L*-curve, is another means of choosing the optimal $\lambda$ (Ref. 48). Solutions to the left of the corner ($\lambda < \lambda_D$) represent 'under-smoothed' solutions, while solutions to the right of the corner ($\lambda > \lambda_D$) represent 'over-smoothed' solutions. The *L*-curve demonstrates that minimizing the residual norm $\rho(\lambda)$ is not an effective means of determining an approximate solution $g_\lambda$. As $\lambda$ is reduced below $\lambda_D$, $\rho(\lambda)$ continues to decrease, but $g_\lambda$ becomes dominated by noise as evidenced by the rapid increase in $\Omega[g_\lambda]$. For a discussion of why the *L*-curve has its shape, and why the optimal $\lambda$ lies in the corner of the *L*-curve, see Ref. 47. It should be observed that $\lambda$ varies fifteen orders of magnitude in Fig. 2, from $10^{-6}$ to $10^9$.

## 3.2. Regularization with noisy data

Since Eq. (3) exhibits an inherent instability for non-exact (noisy) data, it is necessary to test any inversion algorithm in the presence of noise. Figure 3 shows the exact (light curves) and reconstructed (black curves) current profiles for a uniform thin square, where the data $B_z$ has been corrupted with varying amounts of gaussian white noise of variance $\sigma^2 = \alpha \max\{|B_z|\}$, and the GCV method has been used to determine the regularization parameter. For Fig. 3(a) there is no added noise ($\alpha = 0$), and the exact and reconstructed current are in excellent agreement. Note that $\lambda_D > 0$, due to the finite precision of the data. For Fig. 3(b) $\alpha = 0.001$, which is a noise level approximately equal to that typically obtained from the MOI technique[49]. This is a relatively low noise level and results in very good agreement between the exact and reconstructed current as well, but note that $\lambda_{GCV}$ has increased by more than fourteen orders of magnitude relative to the uncorrupted data. Further increases in the added noise lead to larger values of $\lambda_{GCV}$, Fig. 3(c). Figure 3(d) is a 3D plot of noise corrupted data $B_z$ with $\alpha = 0.2$. The signal is barely distinguishable from the noise, but a good representation of the exact current distribution can still be obtained (Figs. 3(e) and 3(f)). While it is unlikely that this extreme level of noise would ever be encountered measuring the magnetic field above a superconductor, it may be common in magnetic inversion problems in other areas, such as medical imaging. It should be emphasized that no knowledge of the exact current distribution was used to obtain the reconstructed current in any of these examples, beyond the assumption of smoothness imposed by $\Omega[g_\lambda]$. The regularized solutions shown in Fig. 3(a,b,c,e, and f) were calculated using $\lambda_{GCV}$, the minimizer of $V_{GCV}(\lambda)$, which depends only on the data and the integral kernel. The minimizer of the true mean square error $\lambda_D$ is also shown in Figs. 3(a,b,c, and f), and with the exception of the uncorrupted data (Fig. 3(a)) $\lambda_{GCV}$ is within ~10% of $\lambda_D$ in each case. As the noise level is varied in Fig. 3, the optimal regularization parameter $\lambda_D$ varies by nearly twenty-four orders of magnitude. The regularized solution $g_\lambda$ is somewhat insensitive to small changes in $\lambda$, and varying $\lambda$ by ~20% or more generally leads to negligible changes in $g_\lambda$. In this respect, Fig. 3 demonstrates that $\lambda_{GCV}$ can be an excellent approximation to $\lambda_D$. For the uncorrupted data of Fig 3(a), the success of the GCV method may appear to be somewhat dubious, since $\lambda_{GCV}$ is nearly five orders of magnitude away from $\lambda_D$. However, the figure clearly shows that the choice of $\lambda_{GCV}$ provides excellent results. For noiseless data, $D(\lambda)$ generally exhibits a very shallow minimum, which results in a large range of values of $\lambda$ (several orders of magnitude) that provide perfectly acceptable results. Davies provides a maximum likelihood method for choosing the optimal value of $\lambda$ that may provide better estimates of $\lambda_D$ in the limit of clean data[31], but otherwise led to under-smoothed solutions in numerical tests.

### 3.3. Comparison to other methods

It is instructive to compare the performance of the present method to other methods under different test conditions. The test conditions include uncorrupted (noiseless) data generated from a homogeneous current distribution and the more practical circumstance of noisy data and an inhomogeneous current distribution. The methods for comparison are the present method, the Fourier-filtering method employed by Roth *et al*[17] and the iterative CG method employed by Wijngaarden *et al*[18] (In Ref. 18 the CG method is referred to as CG-FFT). The latter two methods are among the more successful in the literature and each exhibits a regularizing effect as well.

*3.3.1. Uncorrupted data with a homogeneous current distribution*

The first comparison is made using the homogeneous current distribution of Fig. 3 with uncorrupted (clean) data. Figure 3(a) shows the results for the present method, and they are in excellent agreement. Figure 4 shows the functions $D(\lambda)$, $V_{GCV}(\lambda)$, and $\rho(\lambda)$ for the data of Fig. 3(a), where $\lambda_D$ and $\lambda_{GCV}$ have been marked with open circles. $\rho(\lambda)$ has been normalized by $\left\| B_{nm;z} \right\|_2^2$, and is a strictly increasing function of $\lambda$. The values of $\lambda_D$ and $\lambda_{GCV}$ reveal that GCV may not be able to provide good estimates of $\lambda_D$ in the limit of clean data, but because the minimum of $D(\lambda)$ is extremely shallow in this instance, a very large range of values for $\lambda$ produce equally acceptable results. The minimum of $D(\lambda)$ for this data set is $D(\lambda_D) = 8.1 \times 10^{-8}$, but in this case any value of $\lambda$ that gives $D(\lambda) < 10^{-6}$ produces visually nearly identical results to those presented in Fig. 3(a). Using the criterion $D(\lambda) < 10^{-6}$, any value $10^{-25} < \lambda < \sim 7 \times 10^{-3}$ produces equally acceptable results. This is a range of over 27 orders of magnitude demonstrating that the present method is very insensitive to the value of $\lambda$ for *clean* data.

For the Fourier-filtering method of Roth *et al*[17], a regularizing effect is achieved by low-pass filtering with a Hanning window. For this method the approximate solution $g_{k_c}$ is given by Eq. (11) with the filter $Z_{uv;\lambda}$ replaced by $Z_{uv;k_c}$ where

$$Z_{uv;k_c} = \left(\sqrt{u^2+v^2} < k_c\right)\left(1+\cos\left(\pi\sqrt{u^2+v^2}/k_c\right)\right)/2 \tag{16}$$

and the Boolean notation $(x < y)$ has value 1 if true and 0 if false. Here the cut-off frequency $k_c$ plays the role of the regularization parameter. The normalized mean square error and residual norms for this method may be obtained by replacing $g_\lambda$ with $g_{k_c}$ in Eqs. (13) and (14) resulting in $\rho(\lambda) \to \rho(k_c)$ and $D(\lambda) \to D(k_c)$ respectively. Figure 5(b) plots $D(k_c)$ and the normalized $\rho(k_c)$ function. For $k_c = 0$, $g_{k_c} = 0$ everywhere, and $\rho(k_c)/\|B_{nm;z}\|_2^2 = D(k_c) = 1$. The residual norm $\rho(k_c)$ is a constantly decreasing function of $k_c$, and as $k_c \to \infty$, $Z_{uv;k_c} \to 1$, and $\rho(k_c) \to 0$. The mean square error $D(k_c)$ reaches a minimum value at $k_c = 1604$ (marked with an open circle), which is the optimal regularization parameter in this instance. Using the value of $k_c = 1604$, the exact (light curve) and approximate (black curve) current distributions for this method are shown in Fig. 5(a), and are in excellent agreement. Note that the minimum of $D(k_c)$ is again extremely shallow. Using the same criterion $D(k_c) < 10^{-6}$, any value $135 < k_c < 25000+$ would have provided equally acceptable results.

The iterative CG method also has a well-known regularizing (noise-filtering) effect, and in this method the number of iterations, $k$, acts as the regularizing parameter[29,30]. Note that in this case the regularization parameter takes on only discrete (integer) values. The exact form of the CG algorithm used here can be found in Refs. 18 and 50. With $k$ as the number of iterations, let $g_\lambda \to g_k$, and as before, we define the residual and true mean square error norm for the CG method as $\rho(\lambda) \to \rho(k)$ and $D(\lambda) \to D(k)$ respectively. For the CG method, besides choosing the optimal number of iterations $k$, an initial starting point for $g_k$ ($k = 0$) must be chosen. Using $g_{k=0} = 0$, Fig. 5(d) shows the functions $D(k)$ and the normalized residual norm for successive iterations. In this case, $D(k=0) = \rho(k=0)/\|B_{nm;z}\|_2^2 = 1$, and both functions exhibit a rapid initial decrease. $D(k)$ reaches a minimum at $k = 40725$ iterations, and again the minimum is very shallow. The CG algorithm converged at ~100000 iterations in this example, and further iterations did not change the value of $D(k)$. Using the optimal value $k = 40725$ iterations, the exact (light curve) and approximate (black curve) current distributions for the CG method are shown in Fig. 5(c), and are in excellent agreement. The function $D(k)$ in Fig. 5(d) reaches values several orders of magnitude smaller than $D(k_c)$ or $D(\lambda)$, though visually there is little difference between the solutions of Figs. 3(a), 5(a) and 5(c). $D(k)$ falls below $10^{-6}$ after only 19 iterations, and stopping the iterative CG procedure any time after 19 iterations would have produced visually equivalent results. In Ref. 18, the starting value for $g_{k=0}$ was Eq. (11) with $Z_{uv;\lambda} = 1$. Using this starting value for $g_{k=0}$ the results were nearly the same.

In short, Figs. 3(a), 5(a) and 5(c) demonstrate that all three methods can produce excellent results with uncorrupted (noiseless) data. They also show that all three methods are very insensitive to the choice of their respective parameters when the data is uncorrupted. This insensitivity to the parameter value is one of the problems with testing a method with clean data, since it will be shown that choosing the correct parameter value is more critical with noisy data. Also, the minimum values and the shape of $D(\lambda)$, $D(k)$, and $D(k_c)$ are highly dependent on the precision of the data. All the results presented in Figs. 3, 4, and 5 were computed with 16-digit arithmetic, and with that level of precision the ill-posedness of the problem is scarcely evident, and the CG method clearly produces superior results as measured by the functions $D(\lambda)$, $D(k)$, and $D(k_c)$. When 8-digit arithmetic is used for the same problem, the present method of Regularization is superior, with $D(\lambda)$ able to achieve smaller than values than either $D(k)$ or $D(k_c)$. The minima of all three functions are less shallow using 8-digit arithmetic, though there is still a significant insensitivity to the parameter values. For noisy data, accuracy may be limited to 2-3 digits or less.

*3.3.2  Noisy data with an inhomogeneous current distribution*

While the present Regularization method, the Hanning filter method, and the CG method all perform extremely well with uncorrupted data, any effective comparison of methods must be performed with the more practical case of noisy data and an inhomogeneous current distribution. Figure 6(a) shows the chosen test current distribution while Fig. 6(b) presents the *noisy data* generated from the current distribution that will be used to test the multiple inversion methods. Note that that the data is corrupted with a very small amount of noise ($\sigma^2 = 0.001 \max\{|B_z|\}$), and that the added noise is barely detectable in the image.

Figure 7(a) shows the results of the present method when applied to the data of Fig. 6(b). Current profiles through the center of the sample are shown for the exact (light curve) and approximate (black curve) current distribution, where $\lambda_{GCV}$ was used to calculate the approximate solution. The exact current profile is very well reconstructed, particularly where it is oscillating. Shown in Fig. 7(b) are the GCV function $V_{GCV}(\lambda)$, the true mean square error $D(\lambda)$, and the normalized residual norm $\rho(\lambda)$. The minimums of $V_{GCV}(\lambda)$ and $D(\lambda)$ are marked with open circles. The minimum of $D(\lambda)$ is much sharper in this case, but the GCV method provides excellent results and $\lambda_{GCV}$ is very close to $\lambda_D$. Note that as $\lambda \to 0$, $\rho(\lambda) \to 0$ but $D(\lambda)$ is far from its minimum value. This demonstrates again that minimizing the residual norm $\rho(\lambda)$, and hence finding the solution that best satisfies Eq. (3), is not an effective means for obtaining an approximate solution. Here $\lambda_D$ and $\lambda_{GCV}$ are ~60% of their values for the data of Fig. 3(b), even though the noise level is very similar. The optimal value of λ is not only dependent on the noise present in the data, but also on the shape of the data and hence the shape of the current distribution.

For the approximate solution in Fig. 7(c), the Hanning window method was employed. The mean square error and normalized residual norms are plotted in Fig. 7(d). $D(k_c)$ reaches its minimum value at $k_c = 131$ (marked with an open circle), which is the optimal regularization parameter in this instance. The minimum of $D(k_c)$ is much sharper here than in Fig. 5(b), and the approximate solution is much more sensitive to the choice of $k_c$. As an ad-hoc attempt to choose $k_c$ from the data, let $Z_{uv;\lambda}$ be replaced by $Z_{uv;k_c}$ in Eq. (15) and let $V_{GCV}(\lambda) \rightarrow V_{GCV}(k_c)$. The function $V_{GCV}(k_c)$ is plotted in Fig. 7(d) as well, but the minimum of the ad-hoc GCV function (marked with an open circle) fails to provide an acceptable value of $k_c$. Another means of choosing $k_c$ from the data is the *L*-curve method[47], but this method lead to over-smoothed results in numerical tests. Jooss *et al* have shown that in many cases $k_c$ may simply be chosen empirically.[16] However, here the value $k_c = 131$ from the minimum of $D(k_c)$ was used to calculate the approximate solution (black curve) shown in Fig. 7(c), which is a very good approximation to the exact current profile (light curve). The flat regions of the exact profile are perhaps better recovered than with the present method (Fig. 7(a)), though the oscillatory behavior is less well recovered. Visually the solution may be equally acceptable to the results of the present method, but $D(\lambda_{GCV})$ reached a slightly smaller value than $D(k_c=131)$. This is remarkable, as $\lambda_{GCV}$ was determined automatically using only the data, while $k_c$ was chosen by directly minimizing the mean square error between the approximate and exact solutions (Eq. 14). In this instance, the filter derived from Regularization theory is superior to the Hanning window filter.

The results for the CG method are shown in Fig. 7(e). For this example, the initial $g_k$ was again chosen to be zero everywhere. With $g_{k=0} = 0$, Fig. 7(f) shows the functions D(k) and ρ(k) for successive iterations. D(k) decreases with the number of iterations to a minimum value at k = 8, and then begins to increase. Beyond k = 8, D(k) remains a strictly increasing function of k for at least an additional two thousand iterations. The sharp minimum of D(k) demonstrates a much stronger dependence of the approximate solution on the number of iterations. Initially, the $k^{th}$ iterate $g_k$ approaches the exact solution, but then diverges and becomes dominated by noise. This behavior of the CG method for noisy data is well known and is referred to as *semi-convergence*[29,30]. Due to the semi-convergent nature of the CG method, it is necessary to know when to 'stop' the iterative procedure. This determination can be made empirically,[18] but the GCV method is applicable in this case. Observe that the minimum of D(k) is close to the first minimum of ρ(k). As an approximation to the GCV function for the CG gradient method, Hansen gives

$$V_{GCV}(k) \approx \frac{\rho(k)}{(NM-k)^2}, \tag{17}$$

which is valid when $NM \gg k$ (Ref. 29). When $NM \gg k$, the denominator of Eq. (17) may be weakly stationary, and the first minimum of ρ(k) can provide a good estimate of the optimal number of iterations. In this example, the minimum of D(k) occurred at 8 iterations, and the minimum of ρ(k) at 10 iterations. The difference between D(k=8) and D(k=10) is not large, and Eq. (17) provides an acceptable estimate to the minimizer of D(k). Figure 7(e) shows the exact current profile (solid light curve) and approximate current profiles (dotted and solid black curves) for the CG method with $k = 8$. Employing the method described in section 2.2 above for calculating the derivatives of $g_k$ results in the dotted curve in the Figure. The current profile is dominated by noise, and the exact profile is poorly reproduced. This is due to the large amount of noise that was present in the reconstructed stream function, $g_{k=8}$. Note that for the CG method, the normalized mean square error norm reached a minimum value of only $2.2\times10^{-3}$, whereas the for the present method and the Hanning window method values of $3.8\times10^{-6}$ and $9.8\times10^{-6}$ were obtained respectively. This demonstrates that the CG method had much less of a regularizing effect than the other methods. To compensate for the larger amount of noise, the method used to calculate the derivatives was then changed to have an increased smoothing effect. The black curve of Fig. 7(e) was generated using $n_R = n_L = 5$ for a total of $n_R + n_L + 1 = 11$ data points fit to each quadratic polynomial for the calculation of the derivatives of $g_k$. Further increases in the values of $n_R$ and $n_L$ led to a reduced amount of noise in the flatter regions of the current profile, but the oscillatory behavior of the exact profile became poorly reproduced. Of course, the exact current profile was used to determine the optimal $n_R$ and $n_L$, defeating the purpose of using Eq. (17) to choose $k$. For the application of Eq. (17) the initial guess of $g_{k=0} = 0$ is required. For the proposed starting value for $g_{k=0}$ in Ref. 18, $D(k)$ after one iteration was more than $10^6$ and did not fall below $10^6$ in an additional five thousand iterations. Therefore within five thousand iterations, no acceptable solution was found using Eq. (11) with $Z_{uv;\lambda} = 1$ as the starting value for $g_{k=0}$. A variant of the CG algorithm, CGNE (Ref. 29), provided a superior regularizing effect, achieving a minimum of $D(k)$ of $7\times10^{-5}$ after only 34 iterations. However, this is still inferior to the regularizing effects of the present and Hanning window methods.

*3.4. Other Geometries*

All the examples presented so far have been for thin films using the concept of sheet currents in the fully penetrated state. It is interesting to consider the performance of the techniques in other geometries as well. Figure 8 provides an example of flux screening at a relatively low magnetic field. In this example, the sample is an infinite strip of width 256 μm and thickness 0.3 μm. The magnetic field data was calculated analytically (using the formula present in Ref. 21) at a height of $z = 3$ μm above the sample surface. The inset to the Figure shows the $B_z$ profile at the sample surface, $z = 0$. No noise was added to the data, and the exact current profile is plotted in the figure (solid curve) along with the results for the Hanning window (dotted curve), CG (short-dash curve), and present (long-dash curve) methods. For the present method, $\lambda$ was chosen using GCV, and for the Hanning window and CG methods $k_c$ and $k$ were chosen from the minima of $D(k_c)$ and $D(k)$ respectively. All four profiles are nearly overlapping, and all methods perform equally well in the limit of flux screening with clean data. For noise-corrupted data, performance was similar to that shown in Fig. 7, and $\lambda_{GCV}$ again provided excellent estimates of $\lambda_D$.

In numerical tests with slab geometry, the present method obtained results of quality equal to those for thin film geometry, including the performance of the GCV method. Using slab geometry, Regularization and GCV have previously been applied to determine supercurrents in BiSrCaCuO (BSCCO) tapes.[1,4]

### 3.5 *Influence of the measurement height z*

The degree of ill-posedness of Eq. (3) is controlled in large part by the measurement height, $z$. As $z$ increases, the kernel $K$ becomes smoother and the problem becomes more ill-posed.

Consequently, for increasing $z$, a greater degree of regularization (filtering) will be required, resulting in reduced accuracy and spatial resolution. All three methods explored in this work (Regularization, Hanning window, and CG methods) performed equally well in the clean data limit over a large range of values of $z$. This may appear contrary to the results of Ref. 18, but the comparisons made in that work are not representative of either the present method or the Hanning window method but rather with direct Fourier deconvolution (no regularization) equivalent to $Z_{uv;\lambda} = 1$ in Eq. (11). For noisy data, the present method and the Hanning window method produced similar results (when the optimal $k_c$ was known) at each value of $z$, and the CG method exhibited an insufficient regularizing effect. Figure 9 demonstrates the influence of $z$ on the approximate solution using the present method and the current distribution of Fig. 3. In Fig. 9, $\lambda_D$ (triangles) and $D(\lambda_D)$ (circles) are shown as a function of $z$ for uncorrupted data (bottom two curves) and for noise corrupted data (top two curves) with $\sigma^2 = 0.001 \max\{|B_z|\}$ (as in Fig. 3(b)). For both the uncorrupted and corrupted data it can be seen that $D(\lambda_D)$ is an increasing function of $z$, and hence solution quality is decreasing. Note that noisy data and a small value of $z$ may provide better results than clean data (with 16-digits of precision) and a larger $z$. The behavior of $\lambda_D$ in the plot may appear counter-intuitive; as $z$ increases, the problem becomes more ill-posed and more regularization is required, which would suggest $\lambda_D$ should be an increasing function of $z$. While $\lambda$ controls the trade off between the residual norm $\rho(\lambda)$ and the smoothing norm $\Omega[g_\lambda]$, $z$ has a large influence on the magnitude of $\rho(\lambda)$. This can be understood by noting that $\left\| B_{nm;z} \right\|_2^2$ diminishes rapidly with increasing $z$. Therefore, even though $\lambda_D$ is not an increasing function of $z$ in Fig. 9, the values of $\lambda_D$ do give more weight to the smoothing norm $\Omega[g_\lambda]$ in Eq. (7) as $z$ increases.

In practice, in may be difficult to know accurately the measurement height. In the MOI technique for example, the indicator film itself may be 1-5 μm thick[51,52], so the correct value of $z$ to use may not be clear. Also, the separation between the indicator film and the sample surface in the MOI technique, or sensor to sample distance in the scanning Hall probe method, may be difficult to quantify. Figure 10 examines the effect of error in the value of the measurement height $z$ used for the integral kernel $K$. Using the homogeneous current distribution of Fig. 3, data $B_z$ was generated at a height $z = 5$ μm above the sample surface. No noise was added to the data. For the inversion, 'guess' values $z_g = 1, 3, 5, 5.5$, and 6 μm were used. The present method of Regularization with GCV was used for the inversion. Current profiles through the center of the sample for each value of $z_g$ are shown in the Figure. For $z_g = 5$ μm, the results are the same as that of Fig. 3(a). When the true value of $z$ is underestimated ($z_g < 5$ μm), the value of the current density is generally underestimated and the current distribution appears over-smoothed. When $z$ is overestimated ($z_g > 5$ μm), large spikes occur in the profile at the sample edges and where the current changes sign, and current is observed outside of the sample *opposite* in direction to the current just inside the sample. This suggests a procedure to determine the measurement height $z$. The guess value of $z$ ($z_g$) used in the kernel may be overestimated, and then reduced until the current flowing outside the sample (in direction opposite to the current flowing inside the sample) is just reduced to zero. Johansen *et al* have shown that current may be observed outside of the sample when the $B_z$ data is obtained via the MOI technique, due to errors in $B_z$ caused by the in-plane field effect of the indicator film.[21] Laviano *et al* propose an iterative procedure to correct for this effect.[51] A combination of the iterative procedure of that work, and the procedure described here, may be useful to estimate the effective value of $z$ when the $B_z$ data is obtained via the MOI technique and a good a priori estimate of $z$ is not known.

4. **Discussion**

There is a significant difference in the behavior of the Hanning window, CG, and Regularization methods for uncorrupted and corrupted data. For uncorrupted data, the methods are very insensitive to the choice of their respective parameters, and excellent results can be obtained by all methods. However, the magnetic inverse problem exhibits an inherent instability for *noisy* data, and the ill-posed nature of Eq. (3) is not very apparent when uncorrupted, high precision data is used. For corrupted data, the ill-posed nature of Eq. (3) is clear as evidenced by the behavior of $D(\lambda)$, $D(k_c)$ and $D(k)$ in Figs. 7(b), (d), and (f). The minima of $D(\lambda)$, $D(k_c)$ and $D(k)$ are much sharper, making a good choice of $\lambda$, $k_c$, or $k$ (and hence the degree of regularization) more important. Since any experimental technique for making a spatially resolved $B_z$ measurement (i.e. MOI or Hall probes) exhibits some level of noise, the performance of any method to resolve Eq. (3) should be evaluated under such a noise level, where the ill-posedness of the problem is apparent.

For the present method of Regularization (and for the CG method), GCV provides a remarkable means of choosing the optimal parameter automatically from the $B_z$ data. For the Hanning window method, no automated means of choosing $k_c$ was found, meaning that $k_c$ needs to be determined empirically. In many cases, one has a well defined 'guess' of the true current distribution, and choosing $k_c$ empirically can yield excellent results.[16,51] However, when the underlying current distribution is significantly varying on length scales approaching the spatial resolution of the $B_z$ measurement (as in BSCCO tapes[1,4]), it is this author's experience that in determining $\lambda$ empirically, it can be rather subjective to establish the right balance between spatial resolution and noise filtering. In such an instance, there is a significant advantage to be able to apply the statistical considerations of the GCV method to determine the optimal parameter value. There are limitations to the GCV method, however. It was shown in Figs. 3 and 4 and the GCV may fail in the clean data limit. Also, as GCV is a statistical method, it may also fail in the limit of a small sample size (small number of grid points).[45] In numerical tests, $N_x = N_y = 64$ or more was sufficient get excellent results.

In this work, $J_x$ and $J_y$ were determined by application of Eq. (2) to the approximate stream function, $g$. This method of calculating $J_x$ and $J_y$ was chosen for ease of comparison amongst the different methods. However, it has been shown that $J_x$ and $J_y$ may be determined directly from the data, without first calculating an approximation for $g$.[16,17] For example, for a thin film of finite thickness $a$, $J_x$ can be determined directly by using the integral kernel

$$\widehat{K}_x(u,v) = -i\frac{\mu_0}{2\pi}\left(v + \frac{u^2}{v}\right)\frac{\sinh(\pi a w)}{w^2} e^{-2\pi(z-a/2)w}, \qquad (18)$$

where $w = \sqrt{u^2 + v^2}$, and as before, ˆ denotes a Fourier transform and $z$ is the height of the $B_z$ measurement above the sample surface. Inserting $\widehat{K}_x$ for $\widehat{K}$ in Eq. (9) will yield $J_x$ instead of $g_\lambda$. The GCV method may now be applied (using the kernel $\widehat{K}_x$) to determine the optimal value of $\lambda$ for resolving $J_x$ directly from the data. GCV produces excellent results in this instance as well, and allows all the noise filtering for the current components to be determined through statistical means, rather than applying smoothing polynomials to determine the derivatives of $g$ as described in Sec. 2.2. It can be seen from the integral kernels that resolving $J_x$ from the data is a slightly more ill-posed problem than resolving $g$, and the values of $\lambda_{GCV}$ for each are not expected to be the same, even though the data set ($B_z$) is. Once $J_x$ has been determined, $J_y$ may be found from

$$\hat{J}_y = -\hat{J}_x \frac{u}{v}. \tag{19}$$

Note that determining $J_y$ from $J_x$ (or vice versa) is not an ill-posed problem.

The Regularization theory presented in this study was employed using the DFT. This provides a very simple and computationally efficient implementation of Regularization and GCV. However, the DFT has its implementation issues. The DFT introduces a periodic continuation of the resolved solution ($g$ or $J_x$) that requires $B_z$ to be measured over an area significantly greater than the sample size (about twice the width of the sample). Also, edge effects may give rise to spurious Fourier components. These and other issues of the Fourier method are discussed in more detail in Refs. 16 and 18. Due to these issues, there may be circumstances where it is preferred to implement Regularization theory with matrix inversion methods rather than through Fourier de-convolution. In fact, this is the general case, as only special cases of ill-posed problems (such as convolution equations) offer the opportunity to use Fourier methods. Minimizing the regularization functional of Eq. (7) is equivalent to solving the linear system

$$\left(\mathbf{K}^T\mathbf{K} + \lambda \mathbf{L}^T\mathbf{L}\right)\mathbf{g} = \mathbf{K}^T\mathbf{B}_z \tag{20}$$

for $g$. Here the assumption of the smoothness of $g$ is incorporated through **L**. For the one-dimensional case, the second derivative operator is the tridiagonal matrix

$$\mathbf{L} = \begin{pmatrix} -2 & 1 & & & \\ 1 & \ddots & \ddots & & \\ & \ddots & -2 & 1 & \\ & & 1 & -2 \end{pmatrix}, \qquad (21)$$

which is equivalent to Eq. (8). Eq. (20) is well-posed and may be inverted directly yielding

$$\mathbf{g}_\lambda = \mathbf{A}_\lambda^{-1} \mathbf{K}^T \mathbf{B}_z \qquad (22)$$

where $\mathbf{A}_\lambda = \mathbf{K}^T \mathbf{K} + \lambda \mathbf{L}^T \mathbf{L}$. In this case the preferred form of the GCV function is[29]

$$V_{GCV}(\lambda) = \frac{\|\mathbf{K}\mathbf{g}_\lambda - \mathbf{B}_z\|_2^2}{\text{trace}(\mathbf{I} - \mathbf{K}\mathbf{A}^\#)^2} \qquad (23)$$

where $\mathbf{A}^\# = \mathbf{A}_\lambda^{-1} \mathbf{K}^T$. As an alternative to Eq.(22), one may define $\mathbf{b} = \mathbf{K}^T \mathbf{B}_z$, which allows Eq. 20 to be written as

$$\mathbf{A}_\lambda \mathbf{g} = \mathbf{b}. \qquad (24)$$

The CG method may now be applied to Eq. (24). This requires determination of both $\lambda$ and the stopping index $k$, but it allows the regularizing effects of both methods to be incorporated.[53,54]

Finally, discussion of the speed of the various methods is deserved. The Fast Fourier Transform (FFT) is an algorithm for computing the DFT, and the FFT can certainly be employed where appropriate. Obvious symmetries in $K(x,y)$ may also be exploited to save computation time and storage space, though no attempt to do so was made in this work. It is well known that the time taken to compute the 2D FFT scales as $N^2M^2 \log_2(NM)$ (Ref. 46), but quoting such scaling factors may be misleading. For $N = M = 512$ the total time taken to compute the $N{\times}M$ arrays $g_\lambda$, $J_x$, $J_y$, and $|J|$ from an $N{\times}M$ $B_z$ data array was less than 25 seconds in 16-digit arithmetic on a Sun Blade 100 500 MHz UltraSPARC-IIe coded in FORTRAN. Employing the FFT algorithm, the time taken to compute the DFTs of $K$ and $B_z$, and the multiplication and IDFT required by Eq. (11), was only 22% of the total time taken to resolve $g_\lambda$, $J_x$, $J_y$, and $|J|$ from the data. Only 7% of the total time taken was used to determine the minimum of $V_{GCV}(\lambda)$, and the remainder of the time (71%) was expended through file I/O, calculation of $J_x$, $J_y$, and $|J|$ from $g_\lambda$, and miscellany. The Hanning window method is just as quick if a good value of $k_c$ is known a priori. If $k_c$ needs to be determined empirically, Eq. 3 must be resolved repeatedly for each 'guess' value of $k_c$. In that case, there is a speed advantage to the present method, since $\lambda_{GCV}$ is determined *before* an approximate solution is produced. For the CG method, the calculation of one $N{\times}M$ DFT and one $N{\times}M$ IDFT are required for each iteration, which is significantly slower than the other two methods, though the speed of the CG method (as implemented in Ref. 18) scales in the sample size $NM$ equivalently to FFT methods. In any case, speed should be less of an issue than accuracy.

## 4. Conclusion

In summary, Regularization and GCV have been successfully applied to the problem of resolving 2D currents in superconductors from a 2D magnetic field measurement. The Regularization method produced excellent results over a large range of signal to noise ratios, and the GCV method was highly successful in choosing the regularization parameter automatically and objectively, from statistical considerations. Direct implementation of the CG method produces superior results with high precision data, but here it was not found to have a sufficient regularizing effect for practical noise levels. However, the direct CG method can successfully employ GCV for choosing the stopping index. The Hanning window method exhibits a sufficient regularizing effect for noisy data, producing results nearly equivalent to the present method when a good value of $k_c$ is known. Unfortunately, $k_c$ must be determined empirically at present. The results of this study also show that any method for resolving Eq. (3) should be tested with noisy data.

**Acknowledgements**

Thanks are expressed to G. Wahba (University of Wisconsin – Madison), J. Walker (University of Wisconsin – Eau Claire), G.A. Daniels (University of Wisconsin – Madison) and D.C. Larbalestier (University of Wisconsin – Madison) for useful discussions. This work was supported by the Air Force Office of Scientific Research (AFOSR), the U.S. Department of Energy (DOE), and the AFOSR MURI award.

**References**

\* Electronic address: feldmann@cae.wisc.edu

[1] A. Polyanskii, D. M. Feldmann, S. Patnaik, J. Jiang, X. 5, D. Larbalestier, K. DeMoranville, D. Yu, R. Parrella, IEEE Trans. Appl. Supercond. **11**, 3269 (2001)

[2] A. E. Pashitski, A. Gurevich, A. A. Polyanskii, D. C. Larbalestier, A. Goyal, E. D. Specht, D. M. Kroeger, J. A. DeLuca, J. E. Tkaczyk, Science **275**, 367 (1997)

[3] D.M. Feldmann, D.C. Larbalestier, D.T. Verebelyi, W. Zhang, Q. Li, G.N. Riley, R. Feenstra, A. Goyal, D.F. Lee, M. Paranthaman, D.M. Kroeger, D.K. Christen, Appl. Phys. Lett. **79** 3998 (2001)

[4] D. Larbalestier, A. Gurevich, D. M. Feldmann, A. Polyanskii, Nature **414**, 368 (2001)

[5] X. Y. Cai, A. Polyanskii, Q. Li, G. N. Riley. Jr., D. C. Larbalestier, Nature **392**, 906 (1998)

[6] D.M. Feldmann, J.L. Reeves, A.A. Polyanskii, A. Goyal, R. Feenstra, D.F. Lee, M. Paranthaman, D.M. Kroeger, D.K. Christen, S.E. Babcock, D.C. Larbalestier, IEEE Trans. Appl. Supercond. **11**, 3772 (2001)

[7] D.M. Feldmann, J.L. Reeves, A.A. Polyanskii, G. Kozlowski, R.R. Biggers, R.M. Nekkanti, I. Maartense, M. Tomsic, P. Barnes, C.E. Oberly, T.L. Peterson, S.E. Babcock, D.C. Larbalestier, Appl. Phys. Lett. **77**, 2906 (2000)

[8] A. Palau, T. Puig, X. Obradors, A. Usoskin, H.C. Freyhardt, L. Fernandez, B. Holzapfel, IEEE Trans. Appl. Supercond. **13**, 2599 (2003)

[9] L. Fernandez, B. Holzapfel, F. Schindler, B. de Boer, A. Attenberger, J. Hanisch, L. Schultz, Phys. Rev. B **67**, Art. No. 052503 (2003)

[10] C. Jooss, J. Albrecht, H. Kuhn, S. Leonhardt, H. Kronmuller, Rep. Prog. Phys. **65**, 651 (2002)

[11] A. A. Polyanskii, A. Gurevich, A. E. Pashitski, N. F. Heinig, R. D. Redwing, J. E. Nordman, D. C. Larbalestier, Phys. Rev. B **53**, 8687 (1996)

[12] P.D. Grant, M.W. Denhoff, W. 19, P. Brown, S. Govorkov, J.C. Irwin, B. Heinrich, H. Zhou, A.A. Fife, and A.R. Cragg, Physica C **229**, 289 (1994).

[13] X. Granados, S. Sena, E. Bartolome, A. 8, T. Puig, X. Obradors, M. Carrera, J. Amoros, H. Claus, IEEE Trans. Appl. Supercond. **13**, 3667 (2003)

[14] S. Iliescu, S. Sena, X. Granados, E. Bartolome, T. Puig, X. Obradors, M. Carrera, J. Amoros, S. Krakunovska, T. Habisreuther, IEEE Trans. Appl. Supercond. **13**, 3136 (2003)

[15] A. J. Brook, S.J. Bending, J. Pinto, A. Oral, D. Ritchie, H. Beere, M. Henini, A. Springthorpe, Appl. Phys. Lett. **82**, 3538 (2003); A. J. Brook, S.J. Bending, J. Pinto, A. Oral, D. Ritchie, H. Beere, A. Springthorpe, M. Henini, J. Micromech. Microeng. **13**, 124 (2003).

[16] Ch. Jooss, R. Warthmann, A. Forkl, H. Kronmuller, Physica C **299**, 215 (1998).

[17] B.J. Roth, N.G. Sepulveda, and J.P. Wikswo, Jr., J. Appl. Phys. **65**, 361 (1989).

[18] R.J. Wijngaarden, K. Heeck, H.J.W. Spoelder, R. Surdeanu, R. Griessen, Physica C **295**, 177 (1998)

[19] J. Albrecht, Ch. Jooss, R. Warthmann, A. Forkl and H. Kronmuller, Phys. Rev. B **57**, 10332 (1998).

[20] W. Xing, B. Heinrich, H. Zhou, A.A. Fife, and A.R. Cragg, J. Appl. Phys. **76**, 4244 (1994)

[21] T.H. Johansen, M. Bazilijevich, H. Bratsberg, Y. Galperin, P.E. Lindelof, Y. Shen and P. Vase, Phys. Rev. B **54**, 16264 (1996)

[22] E.H. Rhoderick and E.M. Wilson, Nature **194**, 1167 (1962)

[23] R.J. Wijngaarden, H.J.W. Spoelder, R. Surdeanu and R. Griessen, Phys. Rev. B **54**, 6742 (1996)

[24] G.M. Wing, *A Primer on Integral Equations of the First Kind* (SIAM, Philadelphia, 1991)

[25] J. Hadamard, *Lectures on Cauchy's Problem in linear partial differential equations* (Oxford, London, 1923)

[26] A.N. Tihkonov and V.Y. Arsenin, *Solutions of Ill-Posed Problems* (V.H. Winston & Sons, Washington D.C., 1977)

[27] D.L. Phillips, J. Assoc. Comput. Mach. **9** 84 (1962).

[28] E. Micheli, N. Magnoli, and G.A. Viano, SIAM J. Math. Anal. **29**, 855 (1988).

[29] P.C. Hansen, *Rank-deficient and Discrete Ill-Posed Problems* (SIAM, Philadelphia, 1998)

[30] M. Hanke, *Conjugate Gradient type methods for ill-posed problems* (Longman, Harlow, UK, 1995), Pitman Research Notes in Mathematics 327

[31] A.R. Davies, in *Treatment of Integral Equations by Numerical Methods*, edited by C.T.H. Baker and G.F. Miller (Academic, London, 1982)

[32] G. Wahba, SIAM J. Numer. Anal. **14**, 651 (1977)

[33] G. Wahba, in *Solution Methods for Integral Equations*, edited by M. A. Goldberg (Plenum, New York, 1979)

[34] R.S. Rutman and L.M. Cabral, in *Treatment of Integral Equations by Numerical Methods*, edited by C.T.H. Baker and G.F. Miller (Academic, London, 1982)

[35] A. Tihkonov, Soviet Math. Dokl. **4**, 1035 (1963)

[36] D. Nychka, G. Wahba, S. Goldfarb, and T. Pugh, Journal of the American Statistical Association **79**, 832 (1984)

[37] A.K. Louis, Inverse Problems **8**, 709 (1992)

[38] M. Hanke and P.C. Hansen, Surveys Math. Indust. **3**, 253 (1993).

[39] E.H. Brandt, Phys. Rev. Lett. **74**, 3025 (1995)

[40] C.P. Bean, Reviews of Modern Physics **36**, 31 (1964)

[41] See for example Ref. 26 Chapter 4.

[42] J. Walker (University of Wisconsin – Eau Claire), private communication. The second order accurate central difference approximation for the second derivative $f''$ of a discretely sampled function $f$ is $f''_n = (f_{n-1} - 2f_n + f_{n+1})/\Delta$. Re-writing this expression in terms of the DFT of $f$ gives $f''_n = \frac{-4}{N\Delta^2} \sum_{u=0}^{N-1} \hat{f}_u \sin^2(\pi u/N) e^{-i2\pi u n}$ assuming a periodic extension of the $\{f_n\}$.

[43] V.A. Morozov, Soviet Math. Dokl. **7**, 414 (1966)

[44] The predictive mean square error T($\lambda$) is given by $T(\lambda) = \|(K * g_\lambda)_{nm} - (K * g_{exact})_{nm}\|_2^2$. T($\lambda$) is another potential goodness of fit criterion, but in this paper the choice of D($\lambda$) is used. For the present problem, the minimizers of T($\lambda$) and D($\lambda$) appear to be the same numerically and $\lambda_{GCV}$ can be used to estimate $\lambda_D$. For a discussion of when T($\lambda$) and D($\lambda$) have approximately the same minimizer, see Ref. 45.

[45] G. Wahba, *Spline Models for Observational Data* (SIAM, Philadelphia, 1990).

[46] R.S. Anderssen and P. Bloomfield, Technometrics **16**, 69 (1974); R.S. Anderssen and P. Bloomfield, Numerische Mathematik **22**, 157 (1974)

[47] W.H. Press, S.A. Teukolsky, W.T. Vetterling, B.P. Flannery, *Numerical Recipes in Fortran 77* (Cambridge University, Cambridge, 1992)

[48] See Ref. 29 Section 4.6 and 7.5

[49] D.M. Feldmann, D.C. Larbalestier, Unpublished work.

[50] M. Hestenes, *Conjugate Direction Methods in Optimization* (Springer-Verlag, New York, 1980)

[51] F. Laviano, D. Botta, A. Chiodoni, R. Gerbaldo, G. Ghigo, L. Gozzelino, S. Zannella, E. Mezzetti, Supercond. Sci. Technol. **16**, 71 (2003).

[52] A.A. Polyanskii, D.M. Feldmann, D.C. Larbalestier, *Handbook of Superconducting Matierials* (Bristol, IOP) 1999.

[53] R.H. Chan, J.G. Nagy and R.J. Plemmons, SIAM J. Numer. Anal. **30**, 1740 (1993).

[54] C.C. Paige and M.A. Saunders, ACM Trans. Math. Software **8**, 43 (1982).

**Figure Captions**

Figure 1. The geometry of the magnetic inverse problem. The data $B_z$ is assumed to measured on a rectangular grid of $N \times M$ data points a height $z$ above the surface of the sample with grid spacing of $\Delta_x$ and $\Delta_y$ in the $x$ and $y$ directions respectively. The surface of the sample is parallel to the measurement plane. The sample is of arbitrary shape with uniform thickness $a$, which may be zero when the concept of sheet currents is used or infinite in the case of slab geometry.

Figure 2. *L*-curve demonstrating the trade off between $\Omega[g_\lambda]$ and $\rho(\lambda)$ imposed by the regularization functional. For this example the concept of sheet currents was used, and a 512×512 point grid of $B_z$ data with $\Delta_x = \Delta_y = 1$ µm was generated from a uniform square sample of size 200 x 200 µm at a height of $z = 5$ µm above the sample surface. The data $B_z$ where corrupted with gaussian white noise of variance $\sigma^2 = 0.01 \max\{|B_z|\}$. The parameter $\lambda$ varies from $10^{-6}$ in the upper left portion of the plot to $10^9$ in the lower right. The inset shows the same data on a linear scale. The optimal value of $\lambda$, $\lambda_D$, is marked with an open circle. $\lambda_D$ often lies in the corner of the *L*-curve. Solutions with $\lambda < \lambda_D$ represent under-smoothed solutions, and solutions with $\lambda > \lambda_D$ represent over smoothed solutions.

Figure 3. Reconstructed current distributions for a uniform square thin film with varying amounts of added noise. For this example the concept of sheet currents was used, and a 512×512 point grid of $B_z$ data with $\Delta_x = \Delta_y = 1$ µm was generated from a uniform square sample of size 200×200 µm at a height of $z = 5$ µm above the sample surface. The data $B_z$ were corrupted with gaussian white noise of variance $\sigma^2 = \alpha \max\{|B_z|\}$. a) Linear profile through the center of the sample for both the reconstructed current (black curve) and the exact current (light curve). b) same as a) with $\alpha = 0.001$ c) same as a) with $\alpha = 0.01$ d) A 3D plot of the data $B_z$ after being corrupted with noise with $\alpha = 0.2$. The signal is barely distinguishable from the noise. The linear profile in the inset shows the uncorrupted data. e) The streamlines of the reconstructed current for the noisy data of d). The exact streamlines are uniformly spaced concentric squares. f) same as a) with $\alpha = 0.2$.

Figure 4. The functions $D(\lambda)$, $V_{GCV}(\lambda)$, and $\rho(\lambda)$ for the present method using the uncorrupted data generated from the homogenous current distribution of Fig. 3.

Figure 5. Application of the Hanning window and CG methods for the uncorrupted data of Fig. 3. a) Linear profiles through the center of the sample for the reconstructed current obtained using the Hanning window method (black curve) and for the exact current (light curve). b) The functions $D(k_c)$ and the normalized $\rho(k_c)$ for the uncorrupted data of Fig. 3. c) Linear profiles through the center of the sample for the reconstructed current obtained using the CG method (black curve) and for the exact current (light curve). d) The functions $D(k)$ and the normalized $\rho(k)$ for the uncorrupted data of Fig. 3.

Figure 6. An inhomogeneous current distribution and corresponding noisy $B_z$ data to be used in the comparison of the different inversion methods. For this distribution the concept of sheet currents was used. a) A density plot showing the absolute value of the critical current for the test distribution. The sample is 256×256 μm in size. b) A 3D plot of the *noisy data* for the current

distribution of a). The clean data was generated on a 512×512 grid with $\Delta_x = \Delta_y = 1$ μm at a height of $z = 5$ μm above the sample surface. The clean data was then corrupted with gaussian white noise of variance $\sigma^2 = 0.001 \max\{|B_z|\}$.

Figure 7. Comparison of the present method of Regularization to the Hanning window and iterative CG methods using the noisy data of Fig. 6(b). a) The exact (light curve) and reconstructed (black curve) current profiles for the present method of Regularization. b) The functions $V_{GCV}(\lambda)$, $D(\lambda)$, and the normalized $\rho(\lambda)$ for the Regularization method. c) Same as a) for the Hanning window method. d) Same as b), with respective $V_{GCV}(k_c)$, $D(k_c)$ and $\rho(k_c)$ functions. e) Same as a) for the CG method. The light-dotted and solid-black curves both represent the approximate solution for the CG method, with different methods used for calculation of the derivatives. For the dotted curve, $n_R = n_L = 2$, and for the black curve $n_R = n_L = 5$. f) Same as a) with respective $D(k)$ and $\rho(k)$ functions.

Figure 8. Comparison of the present method of Regularization, the Hanning window method, and the CG method for a case of flux screening at low magnetic fields. The exact (solid black curve) profile is shown, along with the results for the Hanning window (dotted curve), CG (short-dash curve), and present (long-dash curve) methods. The four profiles are nearly overlapping. The $B_z$ data for inversion was calculated analytically on a 512×512 grid at a height of $z = 3$ μm above the sample surface with $\Delta_x = \Delta_y = 1$ μm. The sample was 0.3 μm thick, 256 μm wide, and extended beyond the measurement window in the $\pm x$ directions. The sample $J_c$ was 1 MA/cm$^2$ and the applied field was 0.2 mT. The inset shows the profile of $B_z$ for $z = 0$.

Figure 9. Influence of the measurement height $z$ on $\lambda_D$ and $D(\lambda_D)$. For the bottom two curves, the uncorrupted data of Fig. 3 was used. For the top two curves, the same data was corrupted with gaussian white noise of variance $\sigma^2 = 0.001 \max\{|B_z|\}$. The triangles represent the values of $\lambda_D$ and the circles the values of $D(\lambda_D)$.

Figure 10. The effect of 'guessing' the wrong value of $z$ on the approximate solution. For this example the uncorrupted data of Fig. 3 was used. Current profiles resolved from this data are shown, assuming measurement heights of 1, 3, 5, 5.5, and 6 µm. The exact current distribution is shown as a solid black line, and the exact measurement height is $z = 5$ µm.

**Figure 1**

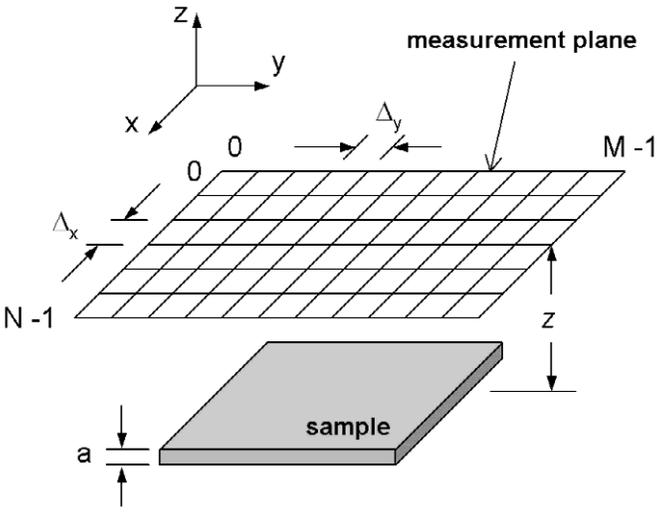

**Figure 2**

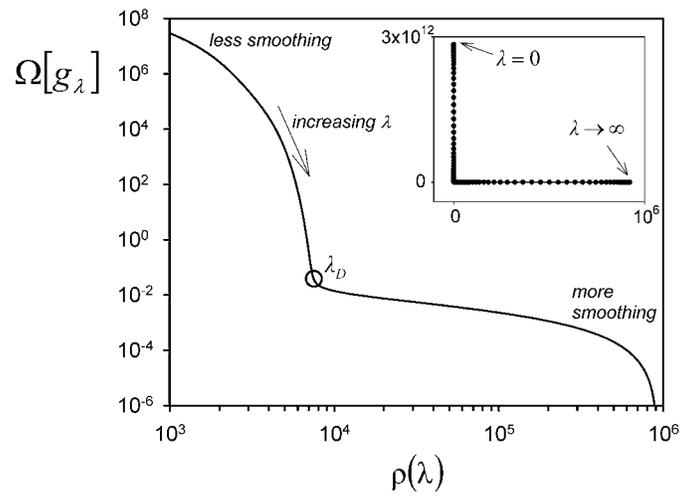

**Figure 3**

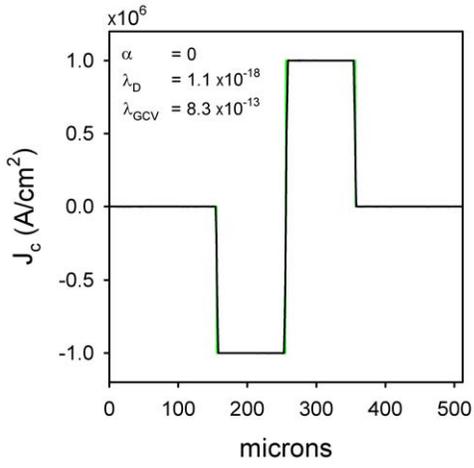
(a)

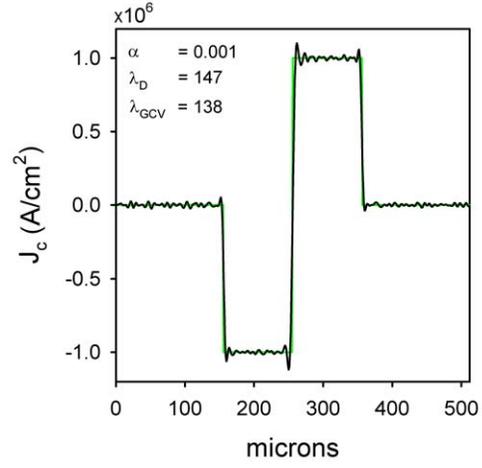
(b)

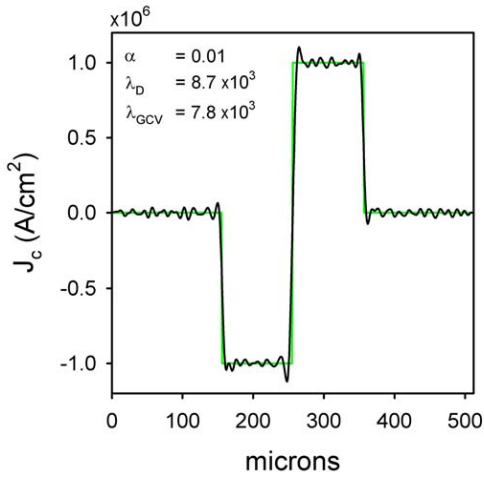
(c)

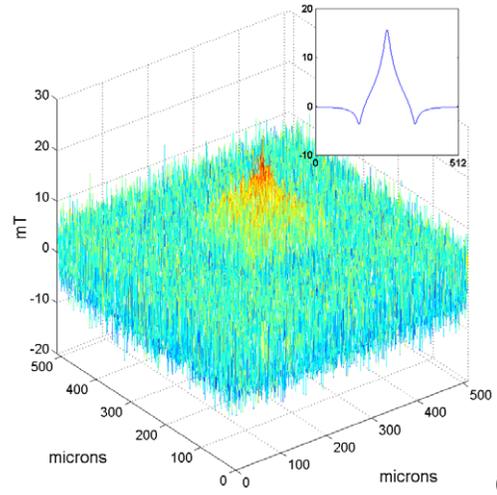
(d)

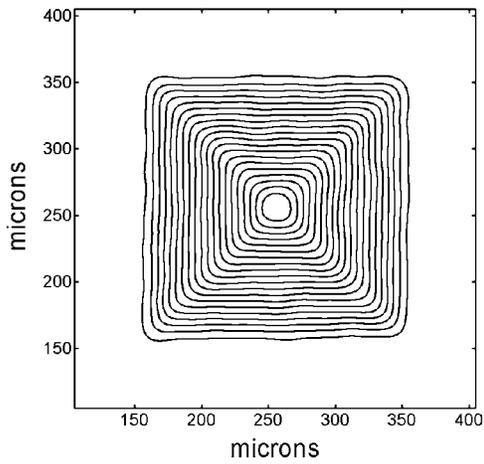
(e)

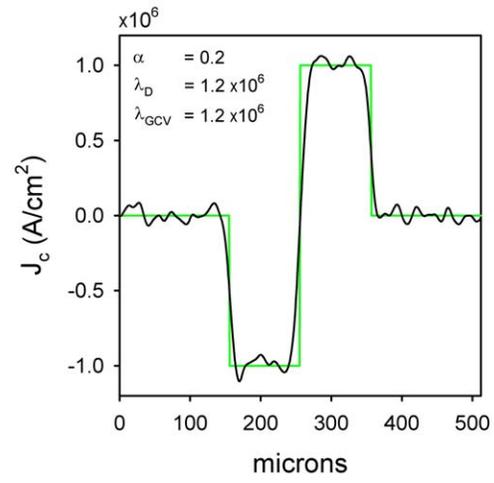
(f)

**Figure 4**

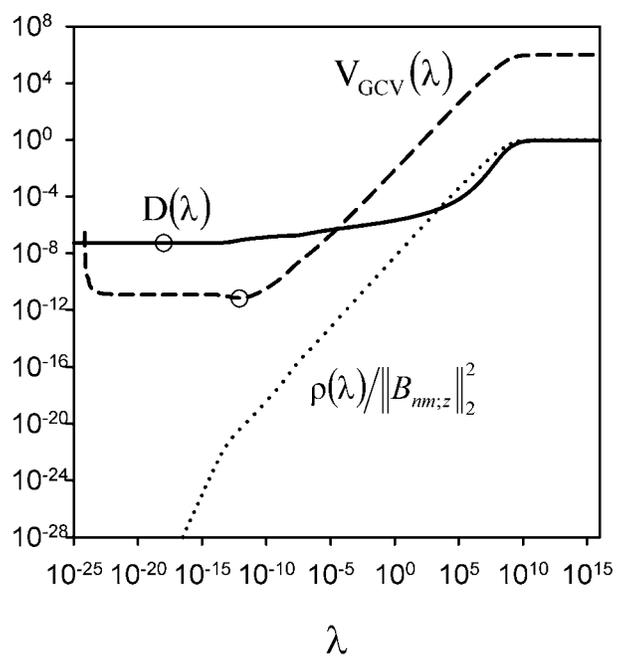

**Figure 5**

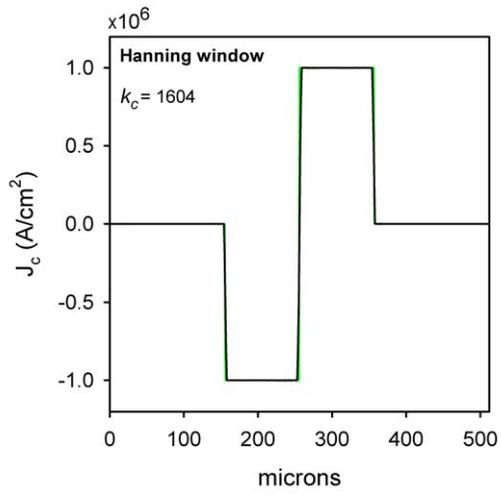

(a)

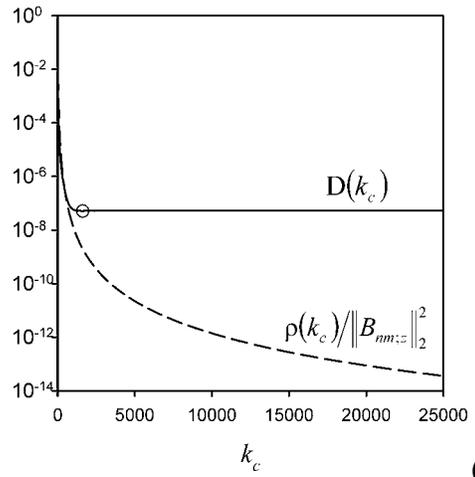

(b)

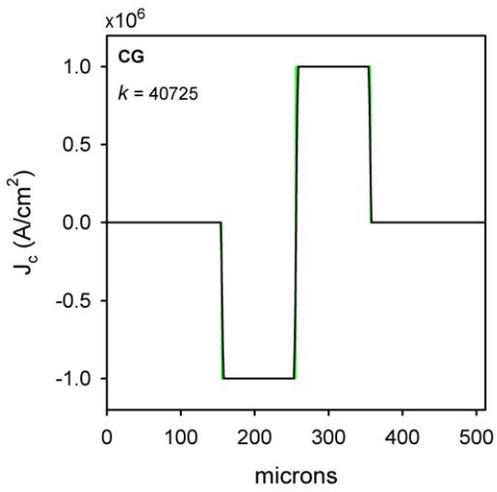

(c)

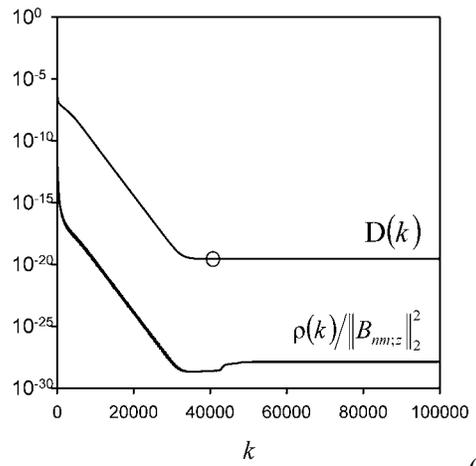

(d)

**Figure 6**

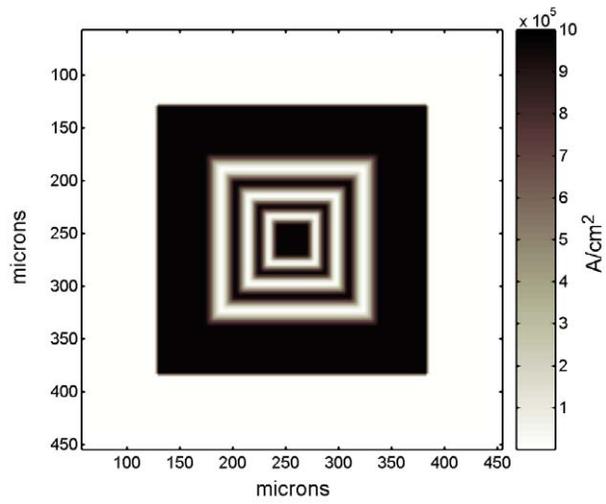 (a)  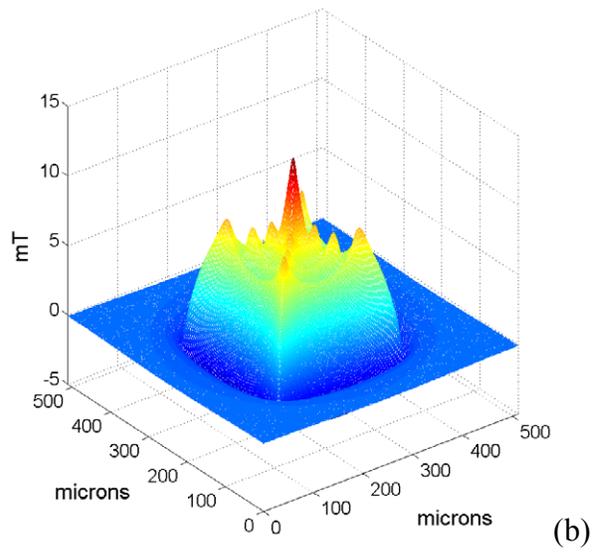 (b)

**Figure 7**

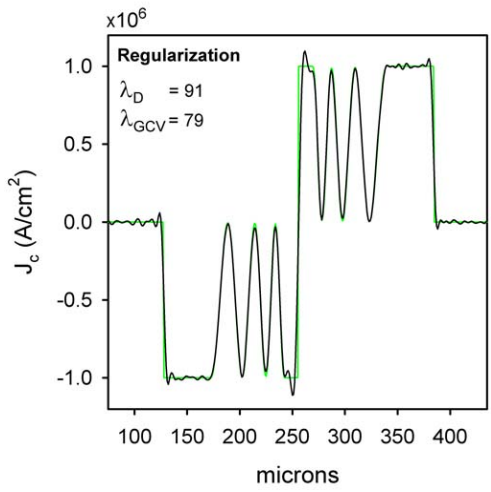 (a)

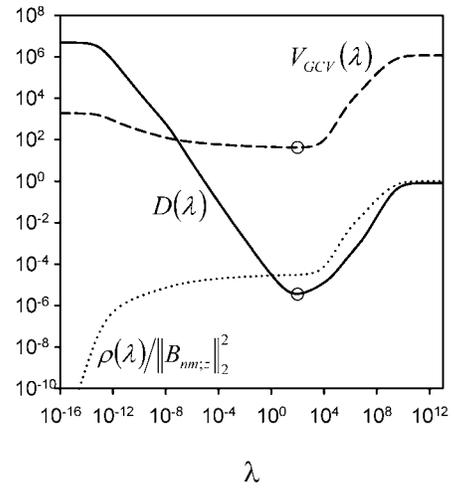 (b)

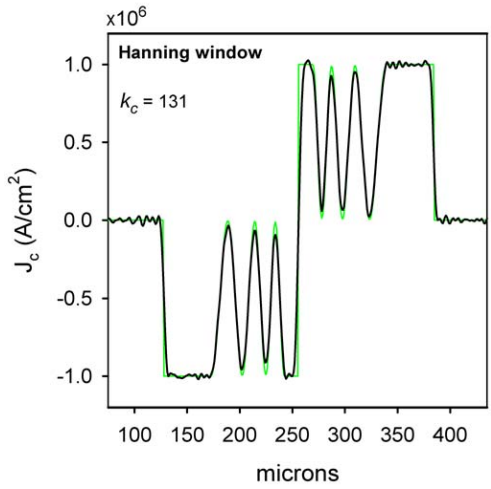 (c)

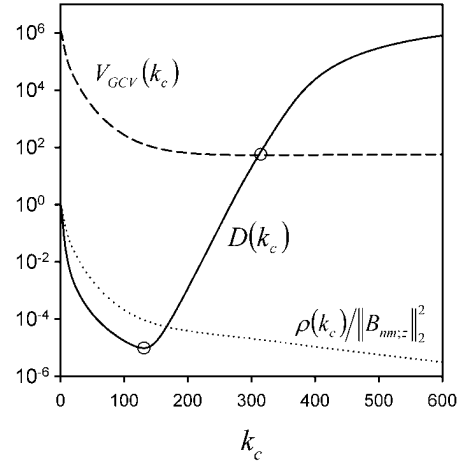 (d)

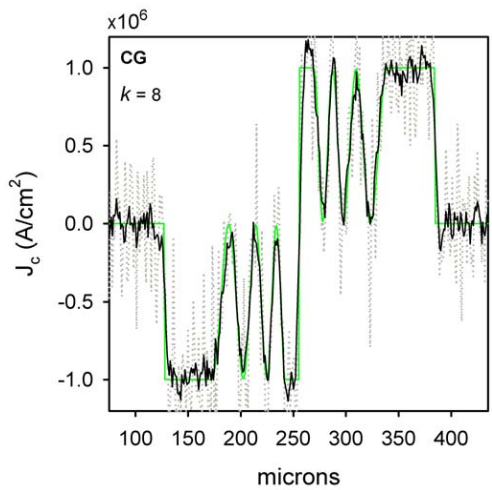 (e)

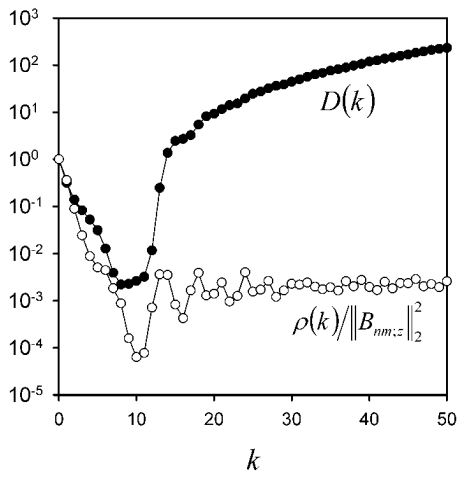 (f)

**Figure 8**

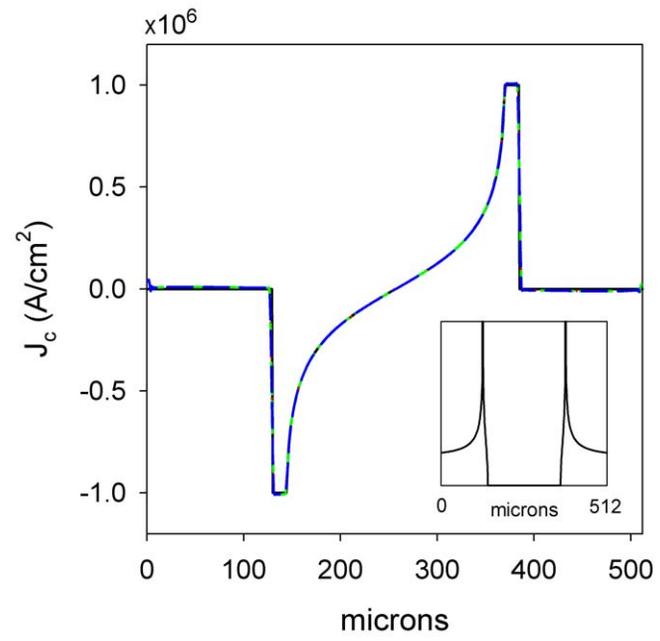

**Figure 9**

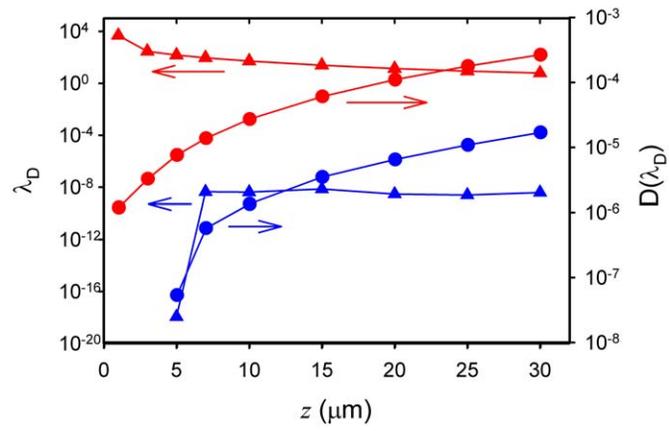

**Figure 10**

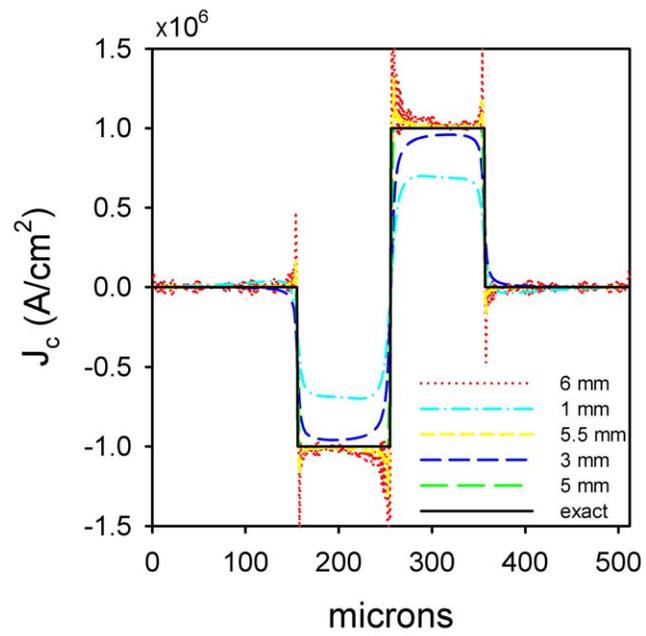